\documentclass[journal]{IEEEtran}

\usepackage{amsmath} 
\usepackage{color}
\usepackage{multirow}
\usepackage{tabularx}
\usepackage[unicode]{hyperref}
\usepackage{cite}
\usepackage{textcomp}
\usepackage[normalem]{ulem}
\usepackage{graphicx}% Include figure files
\graphicspath{ {figures/} }

\newcolumntype{Y}{>{\centering\arraybackslash}X}
\newcommand{\email}[1]{\href{mailto:{#1}}{{#1}}}

\newcommand{\optincludegraphics}[2][]{}

\newcommand{\thejournal}[1]{IEEE MTT}

\title{A 2D-programmable and Scalable Reconfigurable Intelligent Surface Remotely Controlled via\\ Digital Infrared Code}

\begin{document}

\author{
	Andrey~Sayanskiy\textsuperscript{1},
    Andrey~Belov\textsuperscript{1},
	Ruslan~Yafasov\textsuperscript{1},
	Andrey~Lyulyakin\textsuperscript{1},
    Alexander~Sherstobitov\textsuperscript{2},\\
	Stanislav~Glybovski\textsuperscript{1},    
    and Vladimir~Lyashev\textsuperscript{2} \\
    The article has been accepted for publication in IEEE Transactions on Antennas and Propagation.
\thanks{Andrey~Sayanskiy (\email{a.sayanskiy@metalab.ifmo.ru})}
%\thanks{This work has been submitted to the IEEE for possible publication. Copyright may be transferred without notice, after which this version may no longer be accessible}
\thanks{$^{1}$ A.~Sayanskiy, A.~Belov, R.~Yafasov, A.~Lyulyakin and S.~Glybovski are with School of Physics and Engineering, ITMO University, 197101 St. Petersburg, Russia}
\thanks{$^{2}$ A.~Sherstobitov, and V.~Lyashev  are with RTT Algorithm Lab., Huawei Technologies Co. Ltd. Moscow Research Center, 127106 Moscow, Russia}
}

\maketitle

\begin{abstract}

Reconfigurable Intelligent Surfaces (RISs) are promising and relatively low-cost tools for improving signal propagation in wireless communications. An RIS assists a base station in optimizing the channel and maximizing its capacity by dynamically manipulating with reflected field. Typically, RISs are based on dynamically reconfigurable reflectarrays, i.e. two-dimensional arrays of passive patch antennas, individually switchable between two or more reflection phases. The  spatial resolution of provided reflected field patterns is governed by the aperture dimensions and the number of patches to meet the requirements of different communication scenarios and environments. Here we demonstrate a 1-bit RIS for 5-GHz Wi-Fi band made by assembling together multiple independently operating and structurally detached building blocks all powered by the same DC source. Each block contains four separately phase-switchable patch antennas with varactor diodes and a common microcontroller extracting digital control commands from modulated infrared light illuminating the entire RIS. Such distributed light-sensitive controllers grant the possibility of scaling the aperture by adding or removing blocks without re-designing any control circuitry. Moreover, in the proposed RIS a full 2D phase encoding capability is achieved along with a robust remote infrared control.

\end{abstract}

% ======================================================================
\section{Introduction}

Modern and prospective wireless communications systems face with a growing number of simultaneously operating user's terminals and strong interference in urban outdoor and indoor environments. To keep reliable coverage and high channel capacity, conventional fixed antenna systems become insufficient. One of the proposed approaches to improve such systems consists in using passive structures for controlling electromagnetic field distributions. Tunable reflective metasurfaces (MSs - electrically dense and flat periodic structures of subwavelength scatterers) and reflectarrays (RAs - periodic arrays of passive individually tuned antennas with typically a half-wavelength spacing) used for that purpose are commonly called Reconfigurable Intelligent Surfaces (RISs) \cite{DiRenzo_Access_2019,TreryakovIEEE2020,Rui2020,alexandropoulos2020reconfigurable,Tapio2021}. These structures are considered in the literature as powerful and relatively cheap solutions for improving microwave and millimeter-wave wireless channels \cite{Kaina2014,Akyildiz2018}. In outdoor environments, an RIS can assist a fixed base station (BS) dynamically modifying the channel by creating one or several steerable reflected beams as a response to the wave impinging the RIS from the BS. In indoor environments, an RIS can optimize the link by dynamically modifying the spatial distribution of partially standing waves in a given room created by the radiation of a wireless access point, as proposed in \cite{Kaina2014}. In both cases, one or several RISs are to be strategically placed in the environment with a predefined location of the signal source. 

%Applications of RISs, their mathematical models used in wireless communications, as well as their practical realizations have been recently reviewed by several groups of authors. 

%In the microwave range, RISs can be implemented in two ways \cite{Tapio2021}: using MSs or RAs. 

RISs based on MSs consist of deeply subwavelength unit cells and have small periods compared to  $\lambda$, while RAs typically have the period of around $\lambda/2$ ($\lambda$ is the wavelength in free space). In contrast to RAs, MSs operate in the regime of strongly coupled unit cells, so they are modeled as effectively homogeneous impedance boundaries rather than arrangements of discrete elements \cite{glybovski2016metasurfaces,Eleftheriades22}. Despite MSs potentially offer more advanced reflection control such as achieving perfect reflection to angles larger than $70-80^{\circ}$ from the normal \cite{diaz2017generalized}, the features of their unit cells are much finer than of RAs. For that reason, RAs better fit to the relevant technological constraints that are especially important in the millimeter-wave range \cite{Greenerwave2021} and, therefore, serve as RISs in most papers. RAs usually consist of passive metal patch antennas placed over a common ground plane and excited by the incident wave. 

To form a desirable reflected field pattern, the corresponding distribution of the reflection (scattering) phase over the aperture is to be computed and applied to every individual patch.
%The most important function of RISs is to reflect the impinging beam into a desirable direction. 
%For reflection of a normally incident wave to angles of less than $70-80^{\circ}$ from the normal, it is sufficient to approximate a linear reflection phase profile within the aperture. Dynamical steering requires the resonant frequency of each patch is to be gradually tunable. 
The phase can be controlled by changing its resonant frequency of the patch, 
e.g. through biasing a diode connected between the patch and the ground plane \cite{garg2001microstrip} or between two halves of the patch. Gradual and dynamic variation of phase can be provided by varactor diodes \cite{4286010,Sievenpiper2003}, but for a large amount of individually controlled patches it requires complex and expensive electronic circuits. To simplify the design and make it cheaper, 1-bit (also called digital \cite{Cui2014}) RISs have been proposed approximating the phase profile with only two reflection phase states at the expense of high parasitic diffraction lobes in the scattering pattern \cite{7480359}. 1-bit RISs grant a compromising directivity for the main reflection direction provided that the number of patches is high enough (practically, several hundreds and more) \cite{8990007}. Discrete phase variations can be achieved either with varactors \cite{app7090882,Zhang2018,Li_2022} or positive-intrinsic-negative (PIN) diodes. 1-bit RISs switched using PIN diodes have been proposed in \cite{Cui2014,Li2017,Zhang2018NatCom,8993760,Chen:20}. A 2-bit version of an RIS controlled with PIN diodes was presented in \cite{9020088}. While at frequencies below 10 GHz both varactor and PIN diode can be employed, at higher frequencies \cite{9036087}, especially in the 5G FR2 band \cite{8826264,Greenerwave2021,Popov2021} and above \cite{5765450}, PIN diodes have been more popular due to lower losses.
\begin{figure}[!h]
\center
\begin{minipage}{1\linewidth}
\center{\includegraphics[width=0.9\linewidth]{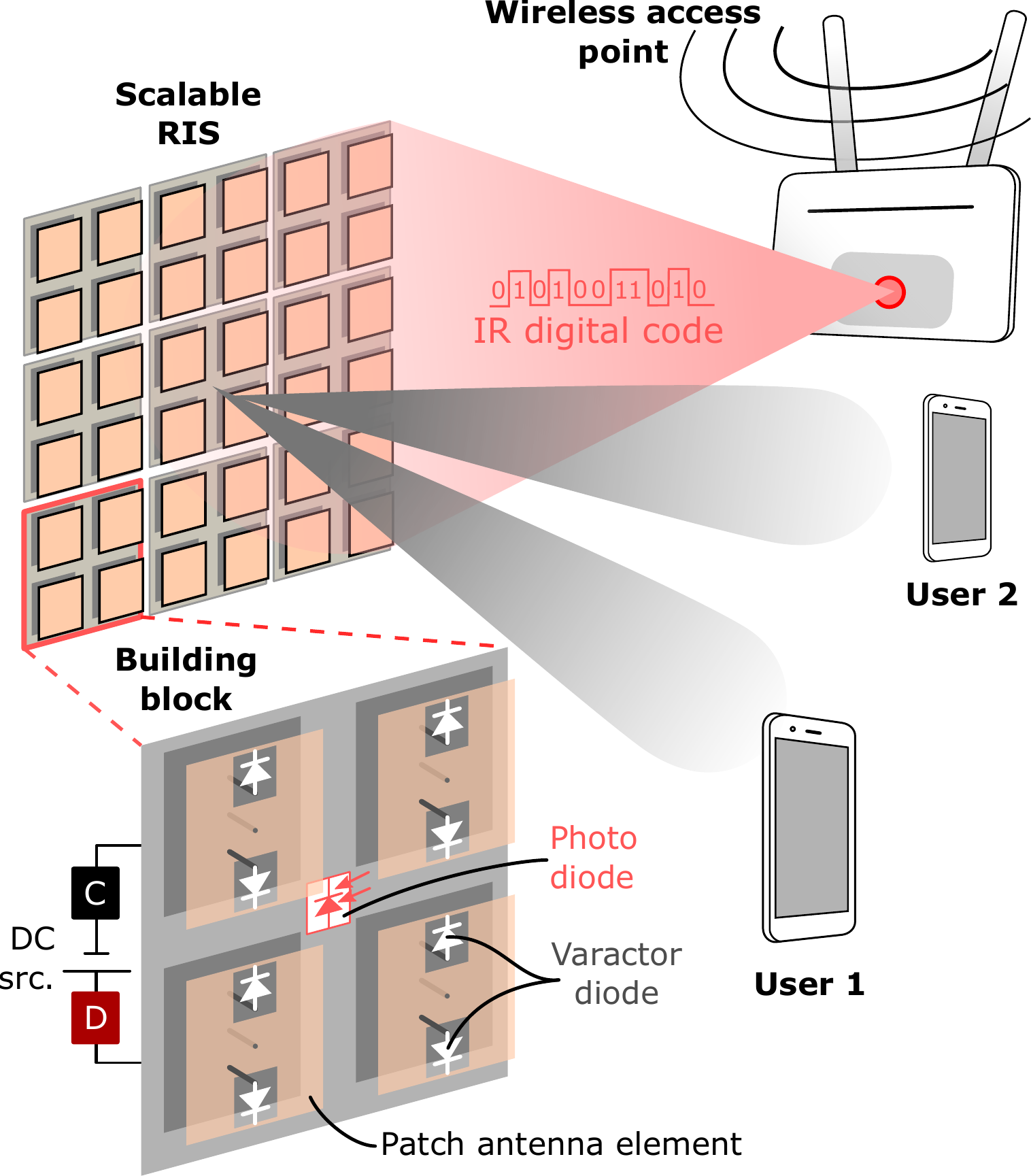}}
\end{minipage}
\caption{Operational principle of the proposed RIS remotely controlled from a wireless access point via infrared digital code. Inset shows the organization of one building block composed of four individually controlled patches loaded with varactor diodes, a photodiode, a controller (C) and DC drive (D).}
\label{fig:Work_princ}
\end{figure}

Special DC circuits are to be embedded into the design of an RIS to individually bias the diodes for setting phases of individual patches (or groups of patches). In order to decouple the DC and microwave parts of the RIS, the DC circuits including biasing lines and controllers are usually placed on the back side of a multi-layer printed-circuit board (PCB). On the contrary, the patches are placed on the front side. For isolation of the two sides, resonant traps (RF chokes) can be used at the corresponding vias \cite{5765450,7509611,Li_2022}. Embedded biasing circuits can contain  distributed controllers placed in every unit cell or usually a group of unit cells. Alternatively, a centralized control unit \cite{Li2017,Zhang2018NatCom,SUN2020883} can be employed. In both cases, changing the aperture dimensions and a number of patches (hereinafter referred to as \textit{scaling}) requires redesigning the entire RIS.
%RISs are to be re-designed each time when the aperture dimensions and a number of patches needs to be changed. %For instance, scaling the aperture is necessary to obtain a desirable width of a reflected beam or certain spatial resolution of the reflected field distribution. 

The  problem of scaling can be solved by using light to deliver bias voltages from a centralized control unit to diodes, or to deliver control signals from a remote control to every element of the distributed controllers. Among others, control with light is the most attractive because reflection phase is encoded in a noncontact manner and with high switching speed \cite{Ren2020}, i.e. without any physical wire connection for transferring control data.

Controlling resonant unit cells in the microwave range with light was first proposed in \cite{Light1} and further demonstrated in electromagnetic structures with various functions e.g. in \cite{Light2,Light2018,Chen:20,Dobrykh2020}. Light-controllable RISs have been implemented in several ways as follows. In \cite{Zhang2018} 1D rows of an RIS were phase-controlled using varactors biased with voltages generated by PIN photodiodes in response to applied visible light of white light-emitting diodes (LEDs). By tuning the intensity of LEDs, the RIS was encoded to produce one or two reflected beams. In \cite{Zhang2020} an RIS with stepwise phase patterns was composed of blocks each one containing $4\times 4$ patches loaded with varactor diodes. The reflection phase of each of $6\times6$ blocks was determined by the intensity of light coming from a meandered chain of LEDs placed behind the RIS. In \cite{SUN2020883} an infrared-controlled RIS capable of 1-bit phase encoding was proposed. Compared to the visible light control, this approach provided longer remote control distances and higher energy efficiency. The RIS had a centralized control unit with several phase patterns pre-coded in FPGA and enabled when receiving coded signals from a remote infrared transmitting circuit. Recently, a scalable RIS with full 2D phase control via dynamically reconfigurable light field illumination from a liquid-crystal display placed behind it has been demonstrated in \cite{Li_2022}. However, to our knowledge, none of the RISs presented in the literature combines full 2D remote control of phase using light with scaling capability.

In this work, we propose and demonstrate a 1-bit RIS with full 2D independent phase control via IR light. In this approach schematically shown in Figure~\ref{fig:Work_princ} IR light coming from a distant source is modulated with digital series control sequences and illuminates the whole aperture of the RIS. The structure is composed of identical and structurally independent building blocks all powered by the same DC source. Each block containing four independently  phase-switchable patches responds only to its own pre-coded address from a sequence and then extracts the information about the required phase states for its four patches. In contrast to the previous approaches, the proposed RIS can be freely scaled by changing the number of blocks without redesigning any circuitry or light sources. 
Distant and robust digital control via IR makes the proposed RIS advantageous for communications inside a room.

This rest of the paper is organized as follows. In Section II, we describe the design procedure for one building block and the entire RIS including the method for extraction of equivalent varactor's parameters. In Section III, the results of experimental investigation are given and compared to simulations and theoretical predictions.

\section{Design and Manufacturing of RIS}

In this Section, we describe the design procedure for one building block of the proposed scalable RIS and of an entire reflective aperture with $20\times20$ individually controllable patches with 2D phase encoding capability operating in the Wi-Fi 5-GHz range. Also, the design of the remote control unit is discussed.

In order to experimentally show the proposed principle the following technical requirements were set: frequency range covering Wi-Fi channels 40, 42 (5.17 - 5.25 GHz), vertical polarization, reflecting aperture dimensions of 600 mm $\times$ 600 mm (approximately $10\lambda\times10\lambda$). %To miminize diffraction lobes of the 1-bit RIS, a square unit cell with dimensions of $\lambda/2 \times \lambda/2$ was chosen with one switchable rectangular patch within. 

Four unit cells were grouped in a building block with one microcontroller separately switching the varactors of four patches between two states. The entire 1-bit RIS was composed of such building blocks all being powered from the same DC voltage.  Surface-mount-device (SMD) varactors SMV2019-040LF by Skyworks Solutions \cite{SMV2019-040LF} were chosen. %switching the patches.

\subsection{RIS building block and unit cell}

The building block with four identical rectangular patches loaded with two varactor diodes each is schematically shown in inset of Figure~\ref{fig:Work_princ}. All four copper patches were printed on the top  of a multilayer PCB with a common ground plane for the microwave and DC parts of the board. The substrate of the patches was made of Rogers 4003C (RO4003) material with a thickness of 1.5 mm, relative permittivity of 3.38 and dielectric loss tangent of 0.0027.
The DC part was organized using multiple additional layers of FR4 behind the ground plane having a total thickness of 1.0 mm, substrate permittivity of 4.3 and dielectric loss tangent of 0.025. The RO4003 and FR4 parts were connected by 0.44-mm-thick prerpeg KB-6065 with a dielectric permittivity of 4.2 and dielectric loss tangent of 0.013. 

\begin{figure}[t]
\center
\begin{minipage}{1\linewidth}
\center{\includegraphics[width=1.0\linewidth]{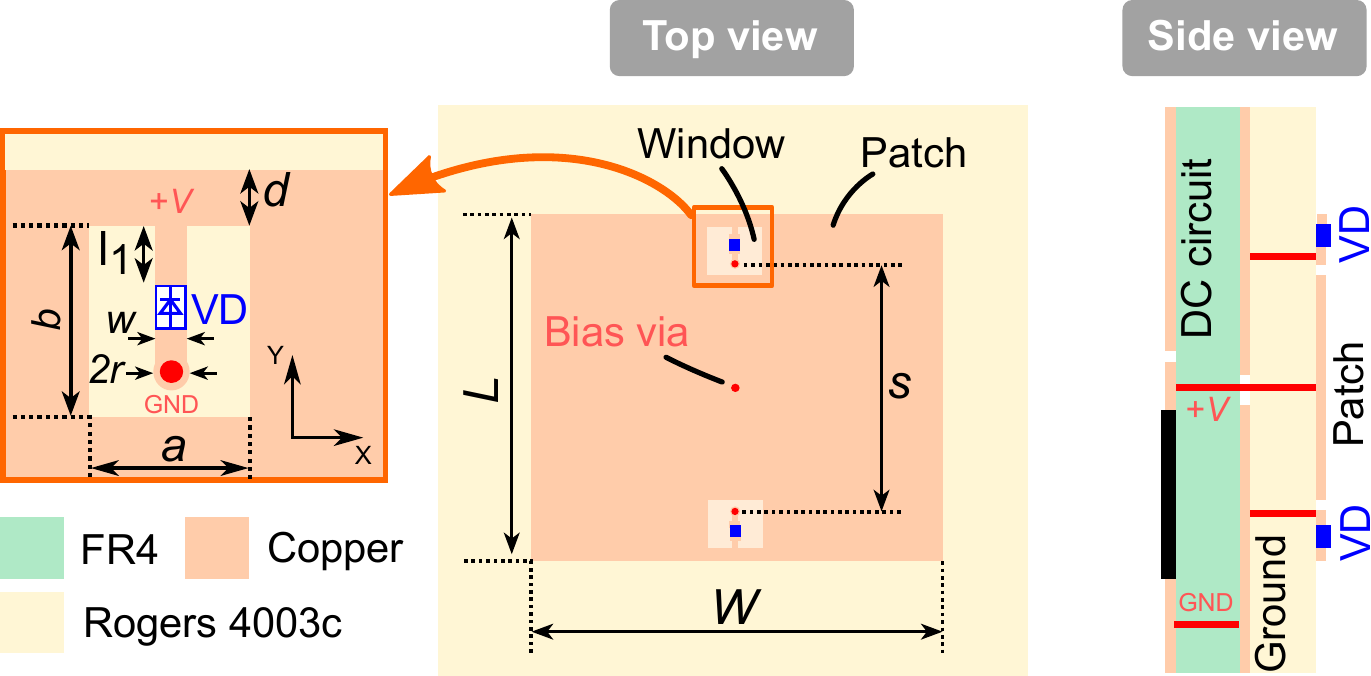}}
\end{minipage}
\caption{Unit cell of the proposed RIS consisting of a rectangular patch on the top layer of a multilayer PCB.}
\label{fig:Unit-cell_geometry}
\end{figure}

To minimize cross-polarization and simplify the interface between the microwave and DC parts of the RIS, we chose a modified version of the classical rectangular patch with dimensions $L$ and $W$ symmetrically loaded with two varactor diodes located near the radiating edges \cite{Bahl} (see Figure~\ref{fig:Unit-cell_geometry}). The patch is assumed to operate with TE$y$ polarization of the incident field having only $y$-component of electric field at any incidence angle. The ground plane was kept at a zero DC potential, while the patch was under a positive bias potential formed by the DC circuit. With this aim, the center of the patch was connected to the DC circuit through a via going through a hole in the ground plane. Also, two additional vias were made to connect the diodes to the ground (marked as GND in the figure). To place the diodes and properly connect them, two rectangular windows with dimensions $a \times b$ were made in the patch and two small contact plates connected to the GNS vias were added. Each diode was soldered between the edge of its window and the corresponding contact plate. For TE$y$ polarization of the incident field, microwave currents are induced only on the patch and ground planes, but no current is induced along the vias. This symmetry allows avoiding RF chokes, which means saving space on the board and reducing losses.

Another advantage of the employed design is the possibility to adjust the effect of the diode's tunable capacitance to the resonant frequency of the patch. The closer the diode and its GND via to the radiating slot, the stronger the reflection phase variation for the same bias voltage. By varying distance $s$ between the windows with varactors it was possible to select two voltages levels corresponding to the 1-bit phase states within an available voltage range of the DC supply for the selected diode. Preferably, the difference between the voltages corresponding to both states should be as large as possible. This condition ensures that the difference in capacitance of the diode switched between two states is considerably larger than any parasitic capacitance due to soldering and mounting on a board. Moreover, that difference is to be larger than a typical capacitance deviation given in the datasheet of the diode. In our case, for stable operation of the unit cell, the difference in capacitance should exceed 0.4--0.5 pF, which was considered when choosing separation $s$.

For the 1-bit operation  we chose "0" state with a reflection phase of $-90^{\circ}$ and "1" state with a reflection phase of $+90^{\circ}$. Both states correspond to equal detuning of the patch resonance from the central operational frequency providing that the magnitudes of the reflection coefficient are equal (with smaller losses than at the resonance). In order to numerically predict the required voltage levels and correctly estimate the losses introduced by the RIS upon reflection, equivalent-circuit parameters of the diode were extracted from waveguide measurements as described below.

\subsection{Extraction of varactor diode's parameters}

\begin{figure*}[t]
\center
\begin{minipage}{1\linewidth}
\center{\includegraphics[width=1.0\linewidth]{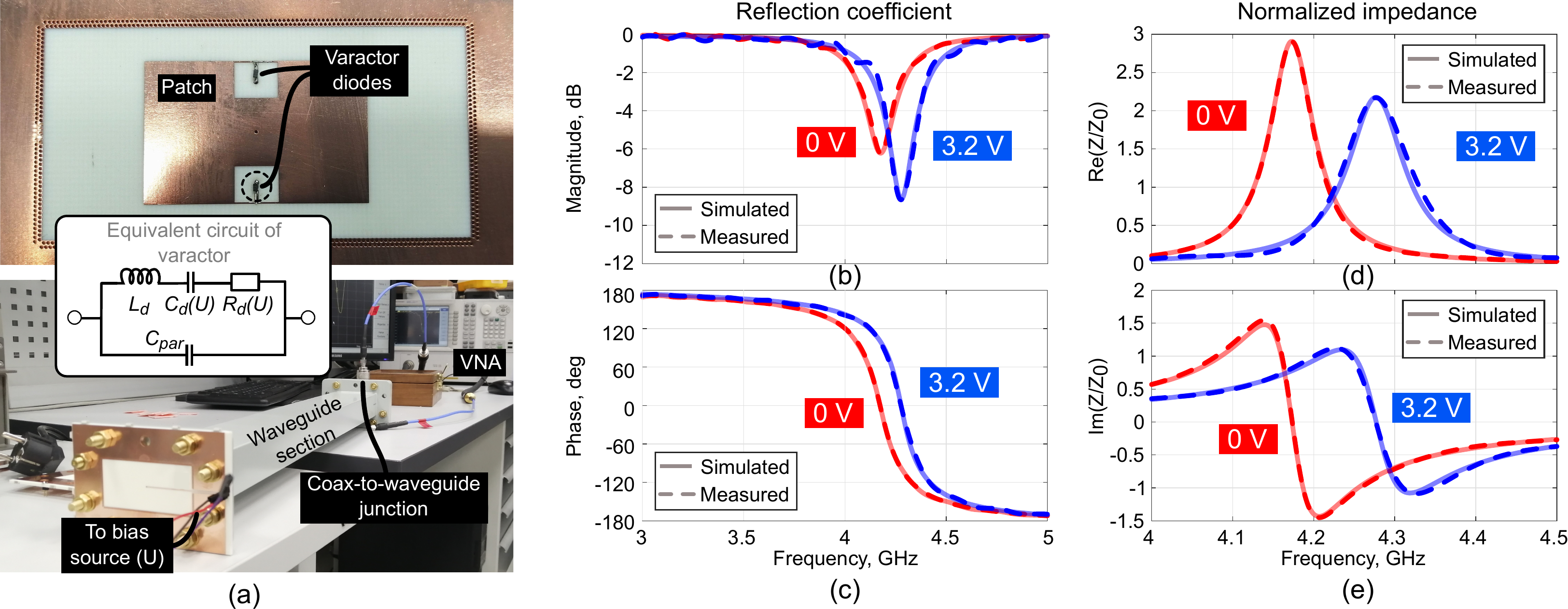}}
\end{minipage}
\caption{To extraction of varactor diode's parameters: manufactured waveguide termination with a patch and two varactor diodes (top) and measurement setup (bottom) with a waveguide section and VNA (a); measured and numerically calculated reflection coefficient magnitude (b) and phase (c); measured and numerically calculated real (d) and imaginary (e) part of the normalized impedance of the waveguide termination. Inset in (a) shows an equivalent circuit used for varactor diodes.}
\label{fig:waveguide_meas}
\end{figure*}

To characterize the varactor diode as a series RLC equivalent circuit we made a set of PCBs serving as resonant waveguide terminations. Like the unit cell of the RIS, each PCB contained a resonant patch with windows and additional contact plates to place two diodes. However, the patch dimensions were changed in order to fit to the cross-section dimensions of WR229 (58.17 mm $\times$ 29.08 mm). With patch dimensions $L=30.5$ mm and $W=19.6$ mm the resonance in the reflection coefficient spectrum was around 4.2 GHz with a zero voltage applied to the diodes (initial estimation was made based on the data from the manufacturer's datasheet). The other parameters according to Figure \ref{fig:Unit-cell_geometry} were: $b=5$ mm, $a=6$ mm, $w=0.5$ mm, $l_1=1.7$ mm, $d=0.2$ mm, $s=13.9$ mm, $2r=0.4$ mm. 
%Three three-layers PCBs were manufactured with one 1.524-mm-thick and  one 0.508-mm-thick Rogers 4003C substrate.
Three different terminations were manufactured each one having three PCB layers: top one with 1.5-mm-thick Rogers 4003C substrate, bottom one with 0.5-mm-thick Rogers 4003C substrate and middle one with 0.2-mm-thick prepreg RO4450 substrate (dielectric permittivity of 3.52 and dielectric loss tangent of 0.004).
The first termination shown in Figure~\ref{fig:waveguide_meas}(a) had two varactors soldered, while the other two terminations used for calibration had no varactors. In one of them the patch was disconnected from the contact plates in the windows, while in the other one short-circuiting metal strips connected the patch to the contact plates instead of the varactors.

The diode's parameters were extracted based on the comparison between the simulated and measured spectra of the reflection coefficient at the input of the waveguide section with a length of 500 mm. All numerical simulations in this work were made in CST Microwave Studio 2020. In simulations both varactors were represented by lumped elements with sizes 0.3 mm $\times$ 0.475 mm. The impedance of the lumped elements was described by the series-type equivalent circuit with capacitance $C_{d}(U)$ and resistance $R_{d}(U)$ depending on bias voltage $U$ as well as parasitic lead inductance $L_{d}$ and parasitic capacitance $C_{\text{par}}$ assumed independent on $U$. The circuit is shown in the inset in Figure~\ref{fig:waveguide_meas}(a).  
The measurements were carried out in the range of 3--5 GHz with a step of 0.5 MHz on vector network analyzer (VNA) Agilent E8362C calibrated using a standard coaxial kit via coaxial-to-waveguide junction (see Figure~\ref{fig:waveguide_meas}(a)). %The PCBs with short-circuiting metal strips and splits at the places of the varactor diodes were used to ensure the correspondence between the calculated and measured  reflection coefficient spectra in the absence of the varactors and to avoid possible numerical errors. 
Parasitic ripples in reflection coefficient spectra caused by multiple reflections in the waveguide section between the sample and coaxial-to waveguide transition were suppressed using the time gating technique. For the PCB with two diodes bias voltage $U$ was varied from 0 to 5 V with a step of 0.2 V.

Two examples of the measured reflection coefficient spectra are given in Figure~\ref{fig:waveguide_meas}(b,c) for $U=0$ V and 3.2 V (the bias voltages eventually selected for the experimental prototype of the RIS). Equivalent-circuit parameters were obtained by their variation in the numerical model to fit the simulated spectra to the measured ones. With this aim, impedance curves of the PCB loads normalized to a characteristic waveguide impedance $Z_0$ were used (see Figures~\ref{fig:waveguide_meas}(d,e)). The frequency of the maximum real part (the resonance), the maximum level for the real part, and the slope of the imaginary part at the resonance were the values to control when fitting.
Note that for different $U$, all three criteria were met only by varying the capacitance $C_{d}$ and resistance $R_{d}$ of the diode. 

The extracted parameters were the following: $L_{d}=0.2$ nH, $C_{\text{par}}=30$ fF. For $U=0$ V the main parameters were  $C_{d}=2.1$ pF and $R_{d}=7.5$ Ohm, while for $U=3.2$ V, $C_{d}=0.87$ pF and $R_{d}=7.1$ Ohm. %Some difference from the equivalent series resistance of 4.5 Ohm given in the datasheet can be explained by the fact that the manufacturer's data was provided for low frequencies. 
The same parameters were further used in optimization of the unit cell of the RIS at 5.2 GHz. 
 
\subsection{Organization and operation of RIS}
 
\begin{figure}
\center
\begin{minipage}{1\linewidth}
\center{\includegraphics[width=0.8\linewidth]{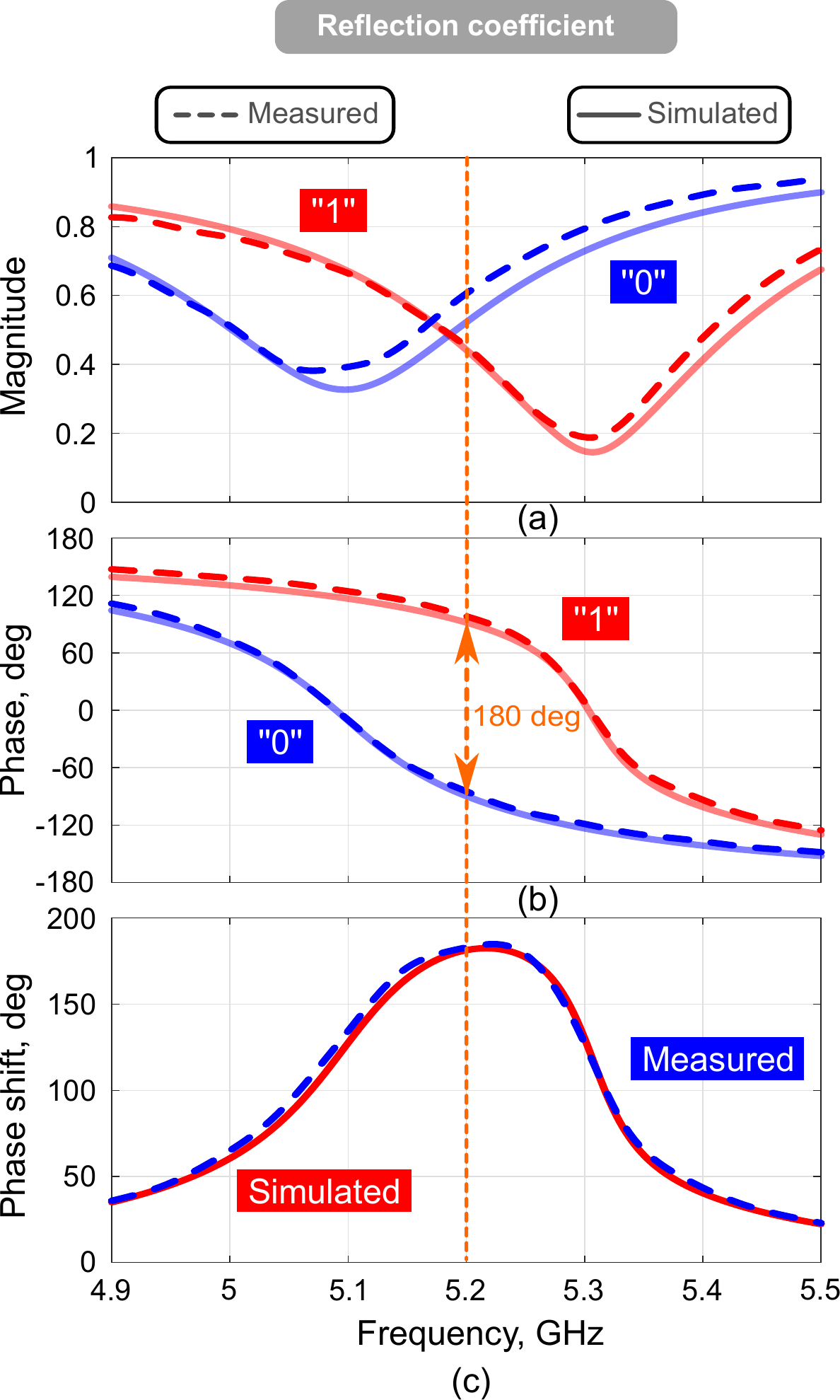}}
\end{minipage}
\caption{Simulated (solid curves) and measured (dashed curves) reflection characteristics of the RIS with all patches in "0" state ($U=0$ V) and in "1" state ($U=3.2$ V): reflection coefficient magnitude (a) and phase (b) for a normally incident vertically-polarized plane wave; phase shift between two reflection states (c).}
\label{fig:Reflection_sim_meas_finite}
\end{figure}

Based on the extracted parameters of the diode, the unit cell was optimized via numerical simulations. The goal was to obtain "0" and "1" phase states at the central frequency of 5.2 GHz for the normal incidence and polarization  along $y$-axis. The unit cell was modeled with a Bloch-Floquet port in conjunction with unit-cell boundary conditions. The resulting parameters (all in mm) read: $L=16.6$, $W=22.0$, $a=2.0$, $b=2.6$, $w=0.475$, $l_1=0.725$, $d=0.5$, $s=11.58$, $2r=0.7$. The reflection coefficient was calculated in the range of 4.9--5.5 GHz (shown with solid lines in Figures~\ref{fig:Reflection_sim_meas_finite} (a-b)). As can be seen, the chosen parameters indeed provide that "0" and "1" states realized with $U=0$ V and $U=3.2$ V, correspondingly, differ by approximately $180^{\circ}$ in the reflection phase. 
%This is additionally confirmed by the plot of the phase difference for the two voltages vs. frequency (Figure~\ref{fig:Reflection_sim_meas_finite}(c)). 
The phase difference holds within the range of 150--210$^{\circ}$ (maximum deviation of phase difference of $30^{\circ}$ from $180^{\circ}$) at frequencies from 5.12 to 5.28 GHz. Note that this operational range covers the target bands of Wi-Fi channels 40 and 42 (5.17-- 5.25 GHz). The reflection coefficient magnitude of around 0.5 in the operational range can be explained by dissipation losses mostly associated with the varactors.

\begin{figure*}
\center
\begin{minipage}{0.8\linewidth}
\center{\includegraphics[width=1.0\linewidth]{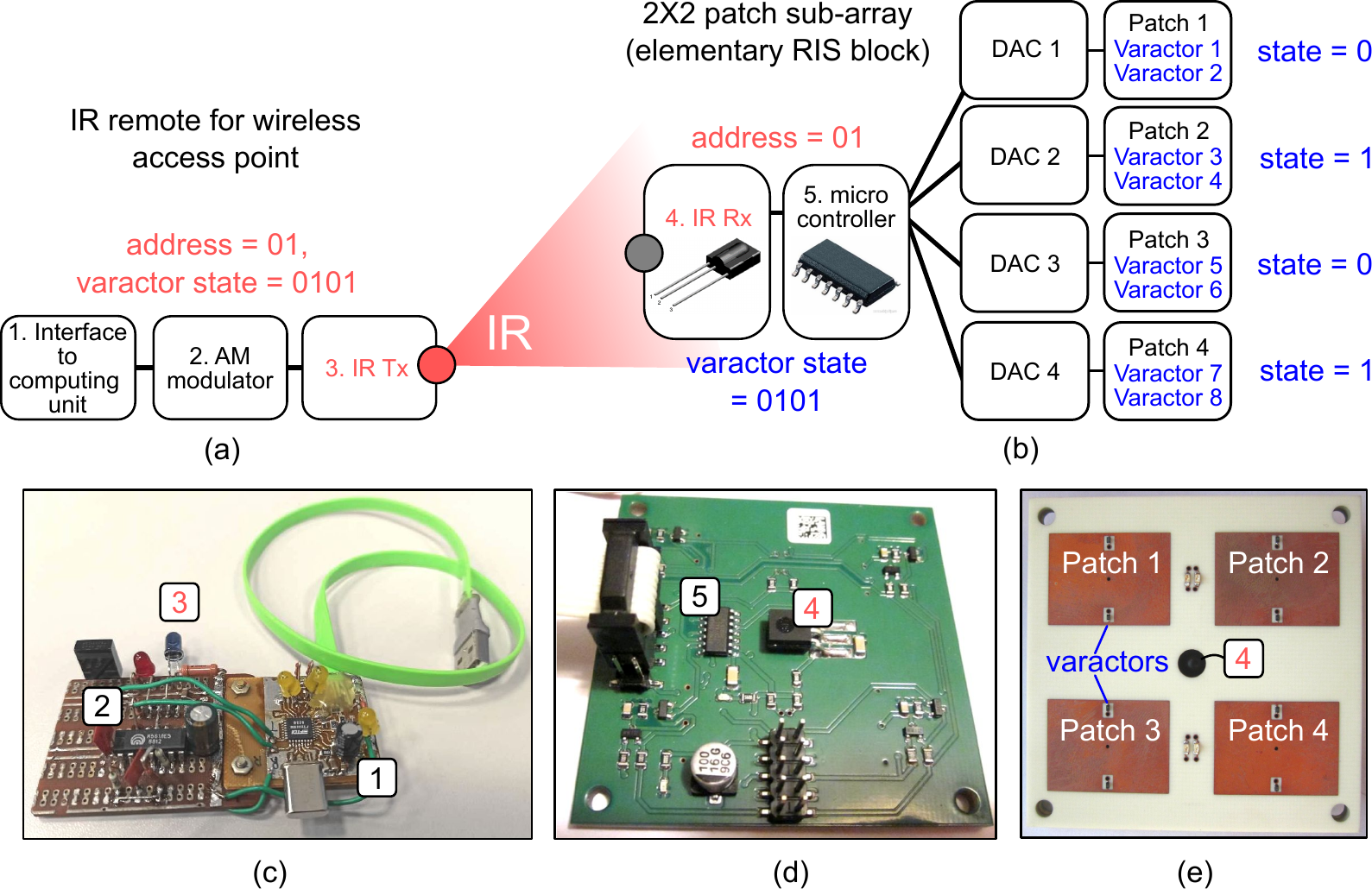}}
\end{minipage}
\caption{Organization of the IR-controllable RIS block and its fabricated prototype: schematic of the IR remote (a) and schematic of the RIS building block with four individually controlled patches (b); prototype of the IR remote circuit with an LED transmitter (c) and prototype of the RIS block shown from the back side (DC biasing circuit elements are shown) (d) and the front side (four patches with a through hole are shown) (e).}
\label{fig:3}
\end{figure*}

Four identical unit cells were combined in one building block having its own control unit that receives control sequences via infrared light regardless its position within the RIS aperture. 
The control sequences were formed by the IR remote control unit. Figure~\ref{fig:3}(a) shows its organisation scheme, where the remote consists of a transmitting near-infrared (940 nm) LED, amplitude modulator and a USB interface to a computer. The IR communication channel used amplitude modulation and control commands produced a series code sequence with the carrier frequency of  38 kHz. The command protocol provided 128 unique addresses (from 0 to 127) which was more than enough to control 100 blocks (i.e. 400 patches).

Within a block, all four patches with four pairs of varactor diodes, and the related components of the biasing circuit were contained within an area of around $\lambda\times\lambda$ on the multilayer PCB having both microwave and DC parts. The biasing circuit contained one TSOP34338 optoelectronic IR receiver module, one ATTiny441 microcontroller, and one driver circuit per patch. 
Figure~\ref{fig:3}(b) shows the scheme of the biasing circuit of a single building block. 
The multilayer PCB of the block had a circular through hole in the center between corners of four patches, which allowed the IR receiver to be sensitive to IR illuminating the face of the RIS. This solution is advantageous over previous light-controlled RISs illuminated from the back as it simplifies mounting low-profile RISs on walls for indoor applications. 

Each block receives IR light illuminating the entire aperture of the RIS and extracts its own control commands from the digital code containing the module unique address and reflection states of four patches. As a response to the remote's commands, the required four bias voltages are set by drivers each one containing a single field-effect transistor (FET), two Shottky diodes and three resistors. Depending on the FET state either one or the other power supply is connected to the driver’s output via a corresponding resistor and Shottky diode. The voltages are then applied to the central vias of four corresponding patches. Four possible voltage combinations are to be stored in the controller. To adjust the voltage levels related to "0" and "1" states for the given varactor diode, each driver only required changing the voltages of two external power supplies.

To test the DC circuit, a prototype of the block was manufactured. Its bottom side (with DC biasing circuit seen) and top side (with four patches and the through hole seen) are shown in Figure~\ref{fig:3} (d) and (e), accordingly. Elements numbering in Figures~\ref{fig:3}(d,e) corresponds to one in  Figure~\ref{fig:3}(b). The manufactured prototype of the IR remote unit is depicted in Figure~\ref{fig:3}(c).

Even one single block being connected to a dual-channel DC power supply may operate as the simplest RIS with only four unit cells. However, to form a practical electrically-large aperture, typically considered in the application, many independently controlled and structurally separated blocks can be assembled together. It is enough to connect all the blocks to the same DC power supply. This scaling principle allows one to build an intelligent surface with a desirable form and size.
Note that adjacent blocks should have their ground planes interconnected. Otherwise gaps between the blocks may affect the operation of the RIS due to resonant diffraction effects. Typically RISs have electrically large aperture dimensions, while RISs with small amount of patches do not make sense. Therefore, it can be recommended to compose the entire RIS from panels each one having at least $5\times 5$ identical proposed blocks. Such large panels can be assembled to a common RIS so that the edges of their PCBs stay close to each other without any electrical connection. The corresponding long gaps are non-resonant and will not affect the RIS operation. In the experiment we used four square $5\lambda\times5\lambda$ panels to compose a RIS with a total amount of $20\times20$ patches.

\subsection{Near- and far-field phase coding}{\label{C}}

The developed RIS belongs to the class of 1-bit reflectarrays with full 2D phase coding capability. To create a desirable spatial distribution of the reflected field, the corresponding phases of individual patches are to be pre-calculated and encoded. The calculation of phases was realized in frames of self-developed software that also controlled the IR remote unit for sending commands to the RIS.

The following adaptive algorithm was developed to calculate the phases based on the specified  distribution of the reflected field. Both the far-field distribution (scattering pattern) and intermediate-field distribution (field pattern on a target plane parallel to the aperture) could be set as the input data. Before starting the adaptation,  the RIS is set to the initial state when all patches are in the same "1" state. However, the algorithm could start from any arbitrary phase distribution. The use of a preliminary approximation usually did not influence on the final result, but could significantly speed up the adaptation process. 

First, the field generated by the RIS in the target area was calculated using a simplified analytical model. The model took into account the excitation field (created by either a plane wave or a point source), the distance from each RIS element to the observation point, the reflection phase of the array element and its individual scattering pattern. All patches were assumed to have a Huygens-like scattering pattern given by the formula: $F(\theta)=\cos^2(\frac{\theta}{2})$ with $\theta$ being a polar angle measured from the normal to the  aperture. If the target area is a plane parallel to the aperture, the reflected field $E_k^{\text{RIS}}$ created by the RIS at point $k$ was calculated as
\begin{equation}
E_k^{\text{RIS}} = \sum_{m=1}^{N_xN_y} E_{\text{inc}}^{m}e^{j\Phi_{\text{inc}}^m} \cdot |R_m| e^{j\Phi_{m}}\cdot e^{-jr_{mk}}\frac{1}{2}\left(1+\frac{z_{\text{plane}}}{r_{mk}}\right),
\end{equation}
where $E_{\text{inc}}^{m}$ and $\Phi_{\text{inc}}^m$ are the magnitude and phase of the incident field at the location of patch element $m$, $|R_m|$ -- magnitude of the local reflection coefficient at patch $m$, $\Phi_{m}$ –- reflection phase of patch $m$ ($-\pi/2$ for state "0" and $+\pi/2$ for state "1"), $r_{mk}$ –- distance between patch $m$ and point $k$, $z_{\text{plane}}$ – distance from the target plane to the aperture of the RIS, $N_x, N_y$ - number of patches along $x$- and $y$-axis respectively. In the case when the source is a plane wave, the phase of the incident field was found as:
\begin{equation}
\Phi_{\text{\text{inc}}}^m = k_0\left[x_{m}\cos(\phi_{\text{inc}})\sin(\theta_{\text{inc}})+y_{m}\sin(\phi_{\text{inc}})\sin(\theta_{\text{inc}})\right],
\end{equation}
where, $x_m$, $y_m$ are Cartesian coordinates of the center of patch element $m$, $\phi_{\text{inc}}$ and $\theta_{\text{inc}}$ -- azimut and polar incidence angles in the spherical coordinate system with a polar axis being normal to the RIS, $k_0=2\pi/\lambda$ -- wavenumber in free space.

Second, the difference between the desired and calculated field distributions was determined. The objective function that needed to be minimized was a mean squared error (MSE) between the two magnitude distributions created on the target plane.

Third, the phase state of one of the patches (selected at random) was flipped and the corresponding change in the objective function was calculated. If the objective function decreased the new phase state was kept, otherwise, the element was returned to its original state and the process repeated.

The criteria to stop the iteration process was the absence of changes in the phase distribution for a specified maximum number of iterations. It should be noted that the number of steps depends on the number of patches and desirable field distribution. For the experimental RIS with $N_x=N_y=20$ it was enough to limit the process with 50 iterations.

The above described objective function can be useful for maximizing the reflected power concentration in the chosen region of space. If this region is small enough, and the distance to the target plane is much larger than the dimensions of the RIS (i.e. $z_{\text{plane}} \gg N_x \lambda/2,N_y \lambda/2$), this process is close to simple beam steering. However, the target region may be large and its dimensions may compare to ones of the aperture. In that case, the process can be applied to holographic field synthesis. In the following section, we consider both types of adaptation made with the experimental RIS.

\section{Experimental results}

\begin{figure}[t]
\center
\begin{minipage}{1\linewidth}
\center{\includegraphics[width=1\linewidth]{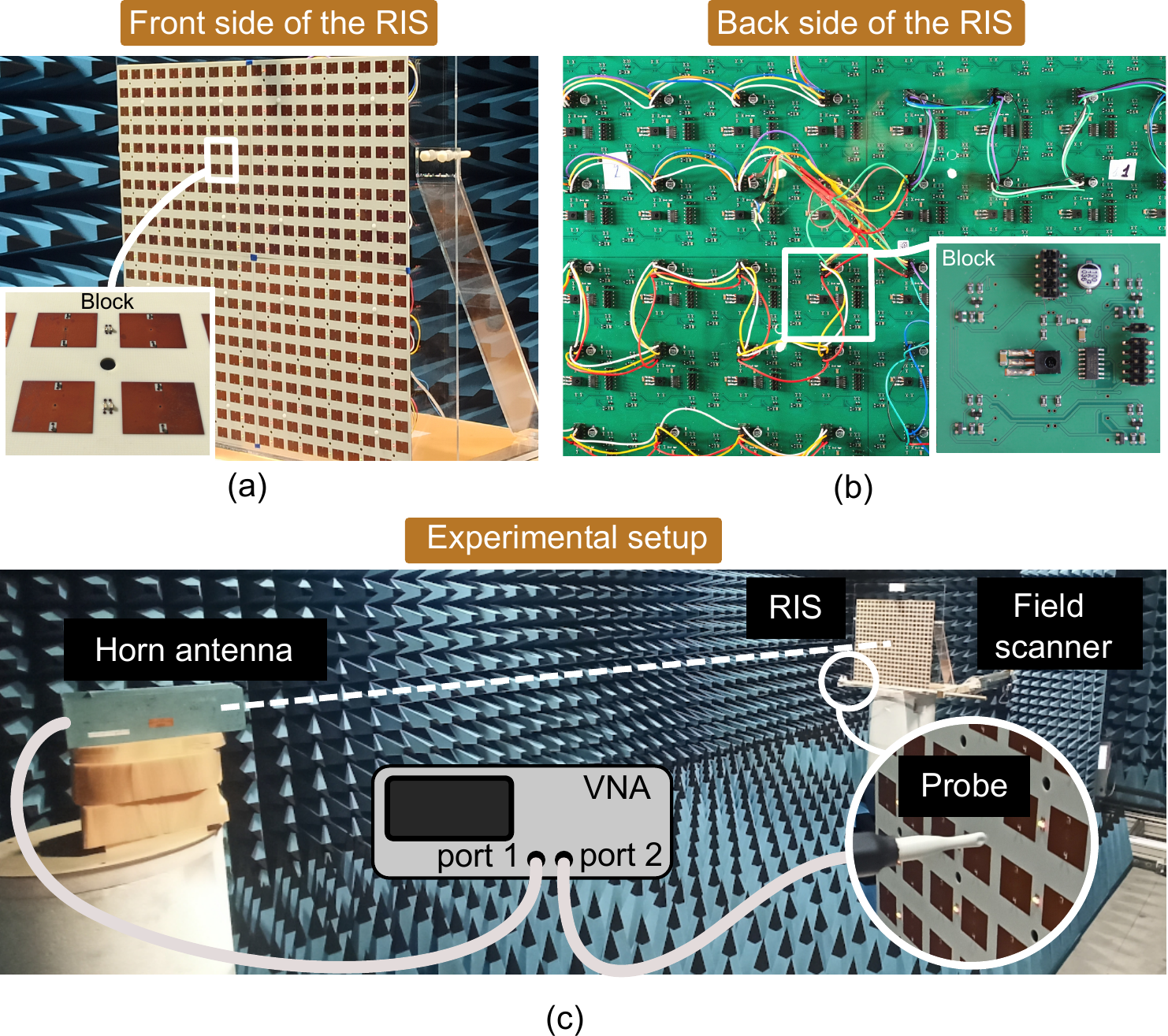}}
\end{minipage}
\caption{(a) Fabricated RIS composed of $N_x\times N_y=20\times 20$ patch elements shown from the front side (inset shows one building block with $2\times2$ patches); (b) back side of the RIS with electronic circuits and wire interconnections for powering building blocks from a common DC supply (inset shows one building block with $2\times2$ patches); (c) Setup for scattering pattern and reflection coefficient measurements}
\label{IRS_photo}
\end{figure}

For the experimental characterization, we fabricated a full-sized prototype of the RIS consisting of $N_x\times N_y=20\times 20$ patch elements (i.e. $10\times10$ IR-controlled building blocks with individual controllers inside). For simplicity of fabrication all blocks were grouped into four identical square panels each one being a separately fabricated multilayer PCB. To support the panels and fix their edges close to each other, a special holder was made of 5-mm-thick Plexiglass. The PCBs were mounted to the holder with plastic screws. The prototype with full dimensions of  600 mm $\times$ 600 mm fixed on the holder is shown in Figure~\ref{IRS_photo}(a) from the front side and in Figure~\ref{IRS_photo}(b) from the back side. DC supply Rohde$\&$Shwarz HMP2030 was used to power the building blocks providing the required voltage levels. 

Measurements of reflected field distributions were made using a 3-axis field scanner installed in an anechoic chamber (7 m $\times$ 5 m $\times$ 3 m) and VNA Rohde$\&$Shwarz ZVB20. The first port of the VNA was connected to the source antenna with a vertical linear polarization ($E_y$) while the second one - to a field probe. 
The experimental setup for the cases of scattering pattern and reflection coefficient measurements is shown in Fig.~\ref{IRS_photo}(c) with the source being a linearly polarized TEM-horn antenna at the distance of 4 m from the RIS.
%The experimental setup for the case of beam steering is shown in Fig.~\ref{IRS_experimental}(c) with the source being a linearly polarized TEM-horn antenna at the distance of 4 meters from the RIS. 
In this case, for obtaining a scattering pattern, first a complex magnitude of total magnetic field $H_x$ was measured over the plane parallel to the RIS at the distance of 10 mm from the patches. Next, the map was remeasured in absence of the RIS (when all four PCB panels together with a Plexiglass holder were removed from the chamber) and was subtracted from the map measured in presence of the RIS to obtain the complex magnitude of the scattered field. Both maps were scanned within the area of 600 mm $\times$ 600 mm with a resolution of 15 mm. Then, Fourier transform was applied to calculate the far-field scattering pattern from the near-field map of the reflected field using standard methods \cite{Kobayashi}. 
In the case of holographic field synthesis, the field was measured directly on the target plane. 
In both cases, SX-R 3-1 H-field Probe mounted to an arm of the scanner and mechanically manipulated in parallel to $XY$-plane was used. The complex magnitude of the measured field was extracted from $S_{12}$ values stored by the VNA.
\begin{figure*}[t]
\center
\begin{minipage}{1\linewidth}
\center{\includegraphics[width=0.85\linewidth]{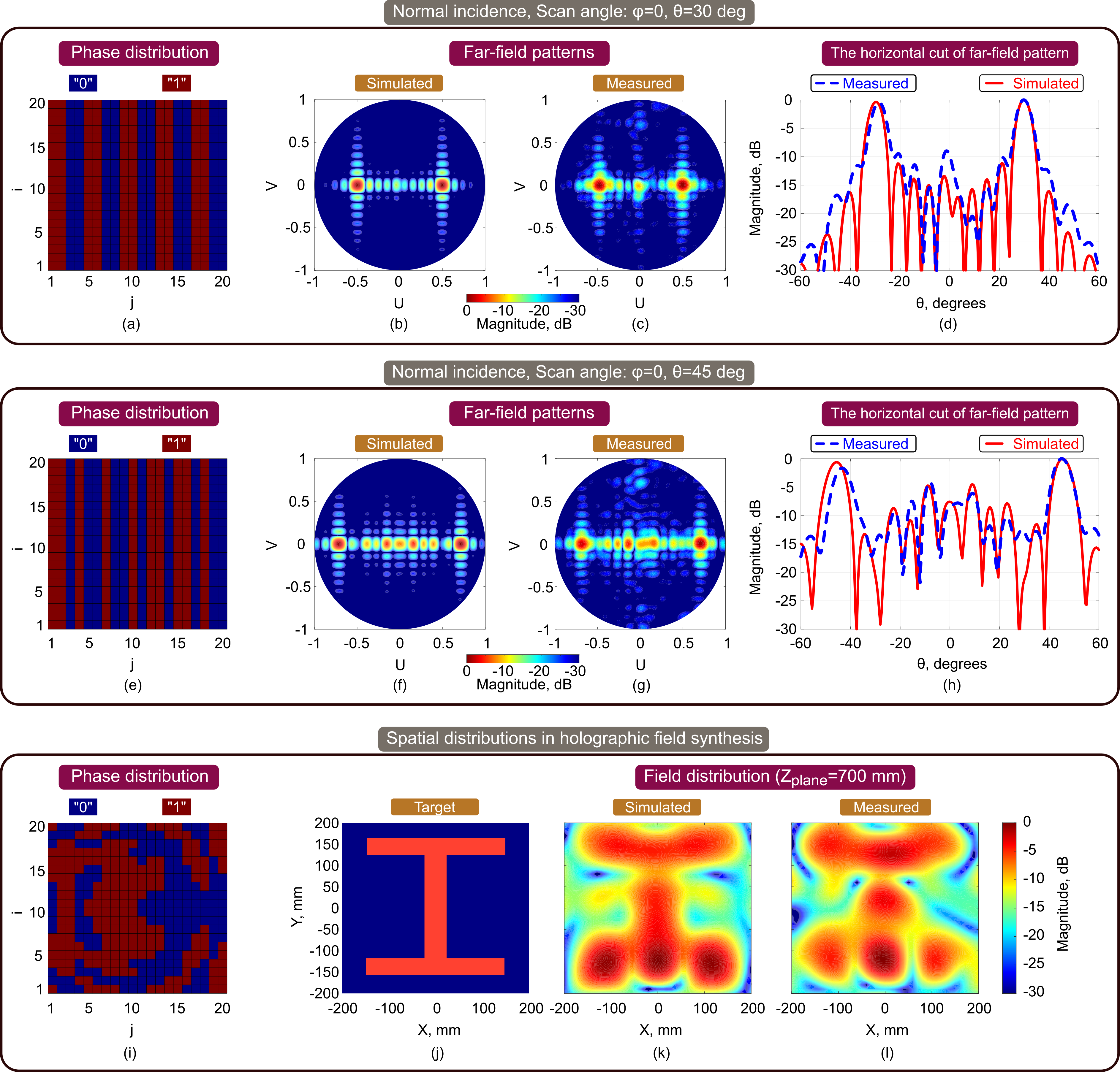}}
\end{minipage}
\caption{Phase distributions with 1-bit coding and corresponding simulated and measured far-field patterns for the normal incidence ($\theta_{\text{inc}}=0$): (a-d) beam steering to scan angle $\theta_{\text{ref}}=30^{\circ}$; and $\theta_{\text{ref}}=45^{\circ}$ (e-h). Phase distribution (i), desirable magnitude field distribution on the target plane $Z_{\text{plane}}=700$ mm away from the aperture (j), simulated (k) and measured (l) field distribution on the same plane.}
\label{IRS_experimental_results}
\end{figure*}

Experimental investigation included: (1) measurements of the reflection coefficients in "0" and "1" applied to all patches; (2) measurements of far-field scattering patterns (regime of reflected beam steering); (3) measurements of intermediate-field distributions in the regime of holographic field synthesis on the target plane positioned at distance $Z_{\text{plane}}=700$ mm away from the RIS; (4) measurements of the signal transmission between two antennas in the presence and absence the RIS.
The results of the corresponding tasks are presented and discussed below in the corresponding subsections.

\subsection{Reflection coefficients}

Magnitudes and phases of the complex reflection coefficient of an incident plane wave from the RIS were measured for various biasing voltage levels applied to the diodes ranging from 2.8 V to 3.6 V with a step of 0.2 V. For each level, the same biasing voltage was applied to all diodes of the RIS (uniform phase distribution). Also the case of a zero biasing voltage was considered in the measurements. The magnitudes and phases were measured indirectly through near-fields \cite{Kobayashi} in the range of 4.9–5.5 GHz with a step of 10 MHz. 
The required near-field maps were scanned within the area of 240 mm $\times$ 240 mm with a resolution of 10 mm. %The indirectly calculated far-field complex magnitudes were obtained at the post-processing step as discussed in \cite{Kobayashi}. 
The scan region was chosen with smaller dimensions than ones of the aperture to reduce the effect of edge diffraction and make the obtained reflection coefficients closer to ones numerically calculated for an infinite 2D-periodic structure. At each probe position the magnitude and phase distributions of the $x$-component of the magnetic field were recorded. The same value in the absence of the RIS and in the case where the RIS was replaced with a copper plate of the same dimensions was also measured to determine a complex reflection coefficient as described e.g., in \cite{Johnson,Sayanskiy}. The effect of parasitic signal reflections between the RIS and the horn antenna was reduced using the time gating procedure \cite{Hock}. 
 
The magnitude and phase of the measured reflection coefficient are compared with the simulation results for $U=0$ V and $U=3.2$ V in Figures~\ref{fig:Reflection_sim_meas_finite} (a) and (b) correspondingly. The phase difference between these two states is shown in Figure~\ref{fig:Reflection_sim_meas_finite}(c). As can be seen, at 5.2 GHz the phase difference between "0" and  "1" states is precisely equal to $180^{\circ}$. Also, the magnitude levels of the reflection coefficient in both these states are almost equal as desired. The measured results are in good agreement with the simulated ones. 

%The scattering pattern was calculated using common post-processing procedure to transform the measured near magnetic field distribution to the far field:
%
%\begin{equation}
%SP^{S}(\phi,\theta) =\sum_{i=1}^{n_x}\sum_{m=1}^{n_y} (S_{12}^{Sample}-S_{12}^{free})cos^2(\frac{\theta}{2})AF\Delta x\Delta y;
%\end{equation}
%
%where, $AF=e^{jk_0(X_{im}cos(\phi)sin(\theta)+Y_{im}sin(\phi)sin(\theta))}$, $\phi$, $\theta$ - angles in spherical coordinate system, $X_{im}$ and $Y_{im}$ - x and y coordinates on the aperture (scan area) respectively, $k_0$ - wavenumber, $\Delta x$ - step along x-axis, $\Delta y$ - step along y-axis.

%The reflection coefficient is calculated with the following expression:
%
%\begin{equation}
%R = \frac{S_{12}^{Sample}-S_{12}^{free}}{S_{12}^{Screen}-S_{12}^{free}}
%\end{equation}
%

\subsection{Scattering patterns in reflected beam steering}

Far-field scattering patterns were measured at 5.2 GHz for the normal incidence, i.e. the wave from the distant horn impinged the RIS from the direction of the normal to its aperture. Depending on  pre-calculated 1-bit phase profiles, different scan angles of an anomalously reflected TE-polarized beam were obtained in the horizontal plane. For comparison, numerical simulations of the entire RIS illuminated with a normally incident plane wave were carried out.

The required 1-bit phase distribution for scan angle $\theta_{\text{ref}}=30^{\circ}$ is shown in Figure~\ref{IRS_experimental_results}(a). A projection of the simulated scattering pattern onto $XY$ plane, i.e. the 2D map of the scattering pattern in dB vs. coordinates $U=\cos(\phi)\sin(\theta)$, $V=\sin(\phi)\sin(\theta)$), is shown in Figure~\ref{IRS_experimental_results}(b). The corresponding measured pattern in shown in Figure~\ref{IRS_experimental_results}(c). The horizontal cuts of the simulated  and measured patterns are shown in Figure~\ref{IRS_experimental_results}(d) in Cartesian coordinates versus angle $\theta$ in the horizontal plane with the red solid line. 
As can be seen from the results, the scattering pattern except for the main beam exhibits a high undesirable symmetric beam. This results in splitting of the reflected power in two beams of almost the same level at $\theta=\pm30^{\circ}$. Appearance of the second beam originates from binary phase quantization and is a common effect of any 1-bit RIS \cite{7480359} in the reflected beam angle differs from the mirror one.
The same data for scan angle $\theta_{\text{ref}}=45^{\circ}$ are given in Figure~\ref{IRS_experimental_results}(e-h).
The simulated and measured maximum directivity as well as the simulated efficiency due to reflection losses in the RIS are given in Table~\ref{T1}.
\begin{table}

\caption{Simulated and measured far-field quantities.}
\label{T1}
\begin{tabular}{lll}
\hline
Quantity    & Simulated  & Measured  \\ \hline
Maximum Directivity, dBi $\theta_{\text{ref}}=30^{\circ}$ & 27.45                &    26.0            \\
Efficiency, $\theta_{\text{ref}}=30^{\circ}$, dB  &      -4.6           &             \\ \hline
Maximum Directivity, dBi $\theta_{\text{ref}}=45^{\circ}$
            &     25.3         &       24.3         \\ 
Efficiency, $\theta_{\text{ref}}=45^{\circ}$, dB
            &     -4.8      &      \\ \hline         
\end{tabular}
\end{table}
%

%
%\begin{table*}
%\textcolor{blue}{
%\caption{\textcolor{blue}{Comparison of our design with previously known approaches}}
%\label{T2}
%\begin{center}
%\begin{tabular}{| c | c | c | c | c | c | c | c |}   %llllllll
%\hline
%Ref. & Control type & Frequency, & Size, & Measured losses, & Measured Directivity, & Number of %photodiodes & Phase control \\ 
%& & GHz & mm & dB & dB & & \\  \hline
%\cite{Zhang2020} & Visible light & 6.5 & 240 $\times$ 240 & no data & no data & 792 & 1D  \\ \hline
%\cite{Zhang2018} & Visible light & 4.02 & 452 $\times$ 452 & no data & no data & 1800 & 1D  \\ %\hline
%\cite{SUN2020883} & IR light & 4.1 & 416 $\times$ 416 & no data & no data & 1 & 1D \\ \hline 
%Our design & IR light & 5.2 & 600 $\times$ 600 & 4.7 & 26 (scan angle: $30^{\circ}$)  & 100 & 2D \\ %\hline 
%\end{tabular}
%\end{center}}
%\end{table*}
%

%
\begin{table*}
\caption{Comparison of our design with previously known approaches}
\label{T2}
\begin{center}
\begin{tabular}{| c | c | c | c | c | c | c | c | c |}   %llllllll
\hline
Ref. & Control type & Frequency, & Size, & Measured & Measured & Number of & Number of & Control \\ 
& & GHz & mm & losses, dB & Directivity, dB & photodiodes & switched cells & source \\  \hline
\cite{Zhang2020} & Visible light & 6.5 & 240 $\times$ 240 & no data & no data & 792 & 36 & 36 LED spotlights \\ \hline
\cite{Zhang2018} & Visible light & 4.02 & 452 $\times$ 452 & no data & no data & 1800 & 36 &  50 white LEDs \\ \hline
\cite{SUN2020883} & IR light & 4.1 & 416 $\times$ 416 & no data & no data & 1 & 64 & 940 nm LED \\ \hline 
Our design & IR light & 5.2 & 600 $\times$ 600 & 4.7 & 26 (scan angle: $30^{\circ}$)  & 100 & 400 & 940 nm LED \\ \hline 
\end{tabular}
\end{center}
\end{table*}

A slight difference between the simulated and measured results is explained by the sphericity of the phase front of the incident wave due to the relatively small distance between the source antenna and the RIS.
Table~\ref{T2} summarizes the features of our design in comparison to previously known approaches. It should be noted that there is a lack of data in the literature on the measured efficiency and directivity RISs. Moreover, none of the RISs presented in the literature combines full 2D remote control of phase using IR remote control with scaling capability.

\subsection{Spatial distributions in  holographic field synthesis}

In order to demonstrate the operation of the 1-bit phase coding in the regime of  holographic field synthesis we aimed to create a certain magnitude distribution of magnetic field on the target plane $Z_{\text{plane}}=700$ mm  away from the RIS. The desirable shape of the distribution was the shape of the capital letter "I" with the sizes of 200 mm $\times$ 200 mm (see Figure~\ref{IRS_experimental_results}(j)).

Using the iterative adaptation algorithm, described in Section \ref{C}, the phase distribution shown in Figure~\ref{IRS_experimental_results}(i) was computed. The simulated and measured H-field distributions on the target plane are shown in Figure~\ref{IRS_experimental_results} (k) and (l) correspondingly. The realized distribution had a blurred shape in comparison with the desirable one due to finite aperture dimensions of the RIS which limited the spatial resolution of the holographic phase synthesis. However, the experimental and simulated distributions resembled shape "I" and looked visually similar to each other.

\subsection{Reflected signal transmission between two antennas}

\begin{figure*}
\center
\begin{minipage}{1\linewidth}
\center{\includegraphics[width=1\linewidth]{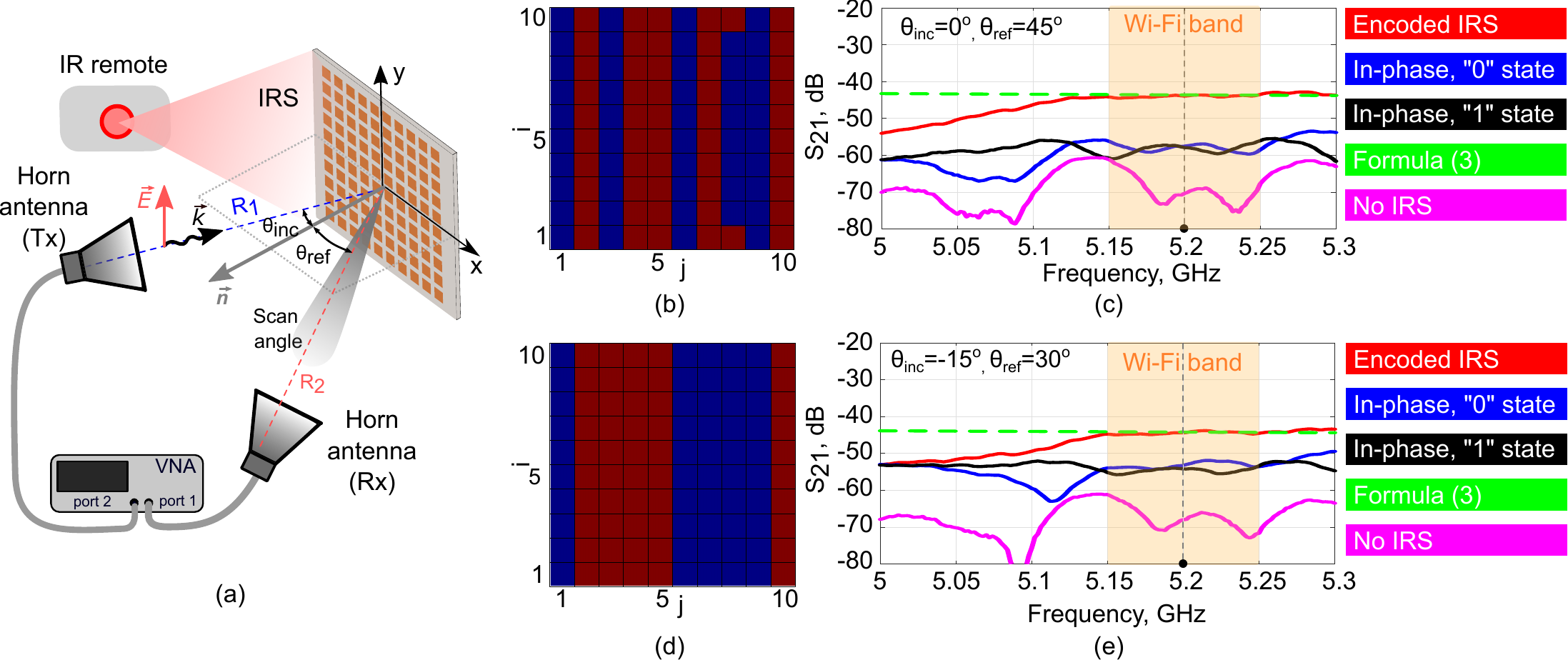}}
\end{minipage}
\caption{To the experimental evaluation of the signal transfer in the presence/absence of the RIS: (a) scheme for measuring the signal transmission between two horn antennas;  1-bit 
phase distribution for the normal incidence ($\theta_{\text{inc}}=0^{\circ}$) and reflection to angle $\theta_{\text{ref}}=45^{\circ}$ (b) and the corresponding magnitude of the transmission coefficient between two horn antennas vs. frequency (c); 1-bit  phase distribution for oblique incidence ($\theta_{\text{inc}}=-15^{\circ}$) and reflection to angle $\theta_{\text{ref}}=30^{\circ}$ (d) and the corresponding magnitude of the transmission coefficient between two horn antennas vs. frequency (e).}
\label{Trans_improvement}
\end{figure*}

The aim of an RIS is to enhance the signal propagation between a base station and a user. To estimate the effect of the fabricated RIS in this practical scenario, we experimentally modeled a single-path wave propagation at 5.2 GHz between two horn antennas with a properly phase-encoded RIS acting as an intermediate reflector on the path. The transmission was characterised by comparing the levels of the signal transferred between the horns in the presence of the RIS and in the absence of it. To satisfy the conditions of the far-field region of the RIS with respect to both horn antennas, in the measurements we used only one panel containing $10\times10$ patches. The experimental setup is schematically  shown in Figure~\ref{Trans_improvement}(a). It consisted of two linearly polarized broadband horn antennas placed at distances $R_1=2.7$ and $R_2=4.2$ m from the RIS. As in previous measurements the horns created the main TE polarization with an electric field vector oriented along $y$-axis. Both horns were connected to ports of VNA Rohde$\&$Shwarz ZVB20 with calibrated 50-Ohm coaxial cables. The signal level was proportional to the magnitude of the $S_{12}$-parameter measured in the range of 5.0--5.3 GHz.
%

%The transmit broadband horn antenna generated a quasi-plane wave with required linear polarization. The antenna was connected to the first port of a vector network analyzer Rohde$\&$shwarz ZVB20 (VNA) by a 50 Ohm coaxial cable. The reflected wave was detected by the receive broadband horn antenna connected to the second port of VNA and $S_{12}$-parameter was stored. For comparison, the same procedure was performed in case of the absence of reflectarray.
%Moreover, the results will be compared with the calculated estimates of the received power by using the radar range equation \cite{Balanis}. %Here we compare the measurements with the estimations made using Friis formula.

%To estimate the efficiency and directivity of the reflectarray and satisfy the requirements to the far-field zone the sub-array with 10x10 array elements was considered. 

The experimental investigation covered several scenarios with different incidence angles and reflection angles in the horizontal ($XZ$) plane realized with different positioning of the horns and RIS in the anechoic chamber. The following scenarios  were compared:
\begin{enumerate}
    \item RIS encoded to reflect a wave coming from a horn at angle $\theta_{\text{inc}}=0^{\circ}$ to a horn at angle $\theta_{\text{ref}}=45^{\circ}$;
    \item RIS encoded to reflect a wave coming from a horn at angle $\theta_{\text{inc}}=-15^{\circ}$ to a horn at angle $\theta_{\text{ref}}=30^{\circ}$;
    \item Mirror reflection with all patches set to "0" state; 
    \item Mirror reflection with all patches set to "1" state. 
\end{enumerate}
The calculated 1-bit phase distributions for the first two (optimized transmission) scenarios are shown in  Figure~\ref{Trans_improvement}(b,d), while the corresponding measured spectra of $S_{21}$ magnitude are given in Figure~\ref{Trans_improvement}(c,e) with red lines. The obtained transmission coefficient levels were found to be in a good comparison to theoretical predictions made using radar range equation (see Section 2.17 in \cite{Balanis}). The received power $P_{\text{r}}$ at horn 2 can be expressed through the transmit power $P_{\text{t}}$ applied to horn 1 as follows:
\begin{equation}
P_{\text{r}} = P_{\text{t}}\frac{\lambda^2L_xL_y\cos(\theta_{\text{inc}})(1-|S_{11}|^2)(1-|S_{22}|^2)}{(4\pi)^3R_1^2R_2^2})G_{\text{t}}G_{\text{r}}G_{\text{RIS}},
\end{equation}
where, $L_x=L_y=300$ mm are the dimensions of the RIS, $\lambda=57.7$ mm - wavelength in free space, $R_1=2.7$ m and $R_2=4.2$ m - distances from horns 1 and 2 to the RIS, $G_{t}=15$ dB - gain of the transmit horn, $G_{r}=15$ dB - gain of the receive horn, $G_{\text{RIS}}$ - gain of the RIS for the given incidence $\theta_{\text{inc}}$ and reflection $\theta_{\text{ref}}$ angles, $|S_{11}|=-20$ dB and $S_{22}=-20$ dB - return loss levels of the transmit and receive horns, respectively. 
It should be noted that this expression is valid only for the case, in which both transmit and receive horns are located in the far-field region of the RIS.
Gain $G_{\text{RIS}}$ was estimated by using the magnitude of the reflection coefficient measurements (0.58 or -4.7 dB at 5.2 GHz) to find the power efficiency and the analytical estimation of the directivity value calculated from the measured scattering pattern. The directivity was estimated as $D=20.1$ dBi in the first scenario and $D=19.5$ dB in the second one. The magnitude of the theoretical transmission coefficient  $|S_{21}|=\sqrt{P_{\text{r}}/P_{\text{t}}}$ is given with green dashed lines in Figures~\ref{Trans_improvement}(c,e). As as can be seen, the theoretical levels well coincide with the measured levels in the first two scenarios. 

For a comparison the measured signal transmission levels are also shown in Figure~\ref{Trans_improvement}(c,e) for the third and fourth scenarios as well as for the case in which the RIS is absent. As follows from the comparison of the results, the presence of the properly phase-encoded RIS improves the signal transmission between the horns by 28.5 and 25.5 dB in the first and second scenario, respectively. In the third and fourth scenario the RIS improves the signal only by less than 12 dB because of a uniform phase distribution and non-optimal mirror reflection behavior.

\section{Summary and Conclusion}

A new approach to design an optically-controlled RIS was proposed and demonstrated. In contrast to previous RISs, structurally and functionally independent building blocks containing four identical patches with an embedded common microcontroller were used. Each block can operate alone or together with multiple similar blocks forming a reflective aperture which is scalable in its dimensions by changing the amount of blocks. Each block can be distantly controlled via IR digital code reacting only to its own commands. The capability of full 2D phase encoding was experimentally demonstrated in the Wi-Fi 5-GHz range in two cases: reflected beam steering in the horizontal plane and holographic field synthesis. Experimental results are in good comparison with analytical predictions and numerical simulations.

The results have shown the possibility of a distant and robust control of a scalable RIS via digital infrared code. Note that during the experiments the IR remote unit equipped with a low-power LED was typically located in a few meters from the RIS provided stable control either in an anechoic chamber or in a furnished lab room. The operational distance of the IR remote is possible to increase by using an IR laser diode or an IR repeater. Optical control is promising since it does not interfere with any radio channels, but limited to using within a single room only. To improve the coverage the proposed scaling principle could be modified to employ other methods of wireless transfer of control signals to distributed controllers, such as via IoT or long-range Bluetooth  protocols.

%In this section, we discuss the obtained results, in particular, the difference between simulations and measurements. Also we revise the obtained technical specifications by the developed IRS.
%We explore the prospective of further scalability of the proposed prototype, and consider the technical issues of going to the higher-frequency range 24-19 GHz.
%The conclusion is made about the proof of principle in the part of scalability of $2\times2$ separately controlled blocks, as well as control signal transmission via infrared light.
%This method eliminates multiple conductors on a printed circuit board (PCB) required to distribute the control signals from a single controller to each array element. It also allows to easily scale the array up and down by removing or adding modules as well as changing the array's shape.

\section{Acknowledgements} 

This work was done in cooperation with Moscow Research Center, Huawei Technologies Co. Ltd., Russia. 
The work was supported in part of waveguide measurements by the Russian Science Foundation (Project No. 21-79-30038).

\bibliography{references}

\end{document}